\RequirePackage{fix-cm}
\documentclass{svjour3}       
\smartqed  
\usepackage{graphicx}
\usepackage{cite}
\usepackage{amsmath,amssymb,amsfonts}
\usepackage{textcomp}
\usepackage{listings}
\usepackage{xcolor}

\usepackage{caption}
\usepackage{multirow} 		
\usepackage{array} 		

\usepackage{url}
 
\usepackage{booktabs} 

\usepackage{makecell}

\usepackage{rotating}
\usepackage[linesnumbered, algoruled, vlined]{algorithm2e}

\SetCommentSty{mycommfont}
\SetKwInput{KwInput}{Input}
\SetKwInput{KwOutput}{Output}

\newcommand{\mygraphic}[1]{\includegraphics[height=#1]{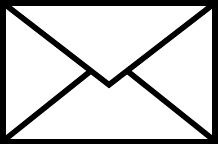}}
\newcommand{\myenv}{(\raisebox{0pt}{\mygraphic{.6em}})}
\newcommand{\myauthor}[1]{#1~\myenv}

\usepackage{ifthen}
\usepackage{amssymb}
\newboolean{showcomments}
\setboolean{showcomments}{true} 
\ifthenelse{\boolean{showcomments}}
{\newcommand{\nb}[2]{
        \fcolorbox{gray}{yellow}{\bfseries\sffamily\scriptsize#1}
        {$\blacktriangleright$#2$\blacktriangleleft$}
    }
    
}
{\newcommand{\nb}[2]{}
    
}

\usepackage{soul} 
\setul{0.5ex}{0.3ex}
\setulcolor{blue}

\newcommand{\etal}{~et al.}					

\lstnewenvironment{jess}[1][]{
	\lstset{
		#1,
		language=Lisp,
		morekeywords={deftemplate,defrule,slot,assert},
		emph={Rule,Node,Sequence,Edge,EbR_entityMapping},
		emphstyle={\color{purple}},
		basicstyle=\footnotesize\ttfamily,
		stringstyle=\color{red},
		keywordstyle=\bfseries\color{blue},
		mathescape=true,
		showstringspaces=false,
		frame=single,
		numbers=left
	}
}{}

\lstnewenvironment{jess2}[1][]{
	\lstset{
		#1,
		language=Lisp,
		morekeywords={deftemplate,defrule,slot,assert},
		emph={Rule,Node,Sequence,Edge,EbR_entityMapping},
		emphstyle={\color{purple}},
		basicstyle=\footnotesize\ttfamily,
		stringstyle=\color{red},
		keywordstyle=\bfseries\color{blue},
		mathescape=true,
		showstringspaces=false,
		frame=single,
		numbers=right
	}
}{}

\begin{document}

\title{A Generic Approach to Detect Design Patterns in Model Transformations Using a String-Matching Algorithm}


\author{
	Chihab eddine Mokaddem
	\and \\
	Houari Sahraoui
	\and
	Eugene Syriani
}


\institute{C. Mokaddem \myauthor{}
	          \and H. Sahraoui \and E. Syriani \at
              Université de Montréal, Montréal, Canada \\
              \email{\{cemo.mokaddem,houari.sahraoui,eugene.syriani\}@umontreal.ca}           
           \and
}

\date{Received: date / Accepted: date}

\maketitle

\begin{abstract}
	Maintaining software artifacts is among the hardest tasks an engineer faces. Like any other piece of code, model transformations developed by engineers are also subject to maintenance. To facilitate the comprehension of programs, software engineers rely on many techniques, such as design pattern detection. Therefore, detecting design patterns in model transformation implementations is of tremendous value for developers. In this paper, we propose a generic technique to detect design patterns and their variations in model transformation implementations automatically. It takes as input a set of model transformation rules and the participants of a model transformation design pattern to find occurrences of the latter in the former. The technique also detects certain kinds of degenerate forms of the pattern, thus indicating potential opportunities to improve the model transformation implementation.
	
\end{abstract}

\section{Introduction}

Model transformation is now the mainstream paradigm to manipulate models in model-driven software engineering (MDE) \cite{Batot2016}.
Designing model transformations is a tedious task. Moreover, like any other code artifact, model transformations evolve and should be maintained. To assist developers in writing and maintaining model transformations, several design patterns have been proposed \cite{Lano2014,Ergin2016}.
These presumably facilitate the comprehension and manipulation of transformation programs \cite{DPDobeidallah}.

Detecting instances of a pattern in a transformation provides valuable information to the developer, such as understanding high-level concepts used, and identifying refactoring and reuse opportunities. This helps, among others, improving the documentation.
However, as for general programs, developers do not always implement perfectly a pattern in model transformations.
Hence, design pattern detection should identify both complete and incomplete occurrences.

Detecting design patterns in model transformations did not get much attention so far from the modeling community. To the best of our knowledge, only the work in \cite{Mokaddem2016} has attempted to automatically detect design patterns in model transformations.
Preliminary results showed that this is an effective technique to find complete and approximate design pattern occurrences.
However, this technique has performance limitations as it relies on a rule inference engine that is time and memory consuming.
Another limitation of this technique is the need to specify a set of detection rules for each pattern.

To find inspiration on how to detect patterns in transformations, we looked at the active community of design pattern detection in object-oriented programs. As reported in \cite{DPDobeidallah}, there are dozens of detection approaches for this family of programs. However, as mentioned in \cite{Gueheneuc2010}, these approaches also suffer from performance problems, because detecting complete and incomplete occurrences is generally costly in time, due to the large search-space that includes all possible combinations of classes.
These approaches are also prone to return many false positives, impeding program comprehension, and cluttering the maintainers' cognitive capabilities.
To address the performance issues, the work by Kaczor \etal{}~\cite{DPDKaczor} uses a string
matching technique inspired by pattern matching algorithms in bioinformatics to identify pattern
occurrences in object-oriented programs. These algorithms allow to efficiently process a large
amount of data if the problem to solve can be encoded as a string matching one.
%

In this paper, we propose a generic technique to detect design pattern occurrences in model transformation implementations, without writing detection code for each design pattern, its variants, and approximations.
Like in Kaczor \etal{}, we rely on a bit-vector algorithm that has proven to be efficient for string matching problems \cite{Myers99}.
The challenge we faced is how to encode model transformations, which are sets of rules linked by control schemes, as strings.
The same challenge arises also in the encoding of the patterns as strings.
We succeed to encode the participants of a patterns as strings, but had to complete our approach by a manual step to combine the identified participant instances to form pattern occurrences.
Thus, the detection consists in an automated step that matches the participant strings of a pattern with rule strings of transformation, and a manual step to complete the occurrences.
In addition to the performance, an advantage of using this approach is the fact both complete and incomplete occurrences can be detected.

We evaluated our approach on a set of 18 transformations.
Our results show that patterns are actually used in transformations and that our approach is able to detect them.
Moreover, we found that patterns are not always used independently, but in combination with other patterns.

The rest of the paper is structured as follows. In Section 2
, we first introduce the basic notions used in our work, and then, discuss the related work.
Section 3 details the different steps of our approach, whereas 
Section 4 describes the different forms of patterns that can be detected by our approach.
We provide an evaluation of the detection approach in  
Section 5.
Finally, we discuss the limits of our approach in 
Section 6, and conclude in 
Section 7.

\bibliographystyle{spmpsci}
\bibliography{paperBibTexDB}

\end{document}